\documentclass[letterpaper,twocolumn,prl,aps,superscriptaddress,floatfix]{revtex4}
\usepackage{mathptmx}

\usepackage[latin9]{inputenc}
\usepackage{amsmath}
\usepackage{amssymb}
\usepackage{graphicx}
\usepackage{esint}
\usepackage{float}
\usepackage{array}
\usepackage{ulem}

\makeatletter

\pdfpageheight\paperheight
\pdfpagewidth\paperwidth

\providecommand{\tabularnewline}{\\}

\@ifundefined{textcolor}{}
{%
 \definecolor{BLACK}{gray}{0}
 \definecolor{WHITE}{gray}{1}
 \definecolor{RED}{rgb}{1,0,0}
 \definecolor{GREEN}{rgb}{0,1,0}
 \definecolor{BLUE}{rgb}{0,0,1}
 \definecolor{CYAN}{cmyk}{1,0,0,0}
 \definecolor{MAGENTA}{cmyk}{0,1,0,0}
 \definecolor{YELLOW}{cmyk}{0,0,1,0}
}

\usepackage{xcolor}\usepackage{soul}

\newcommand{\ket}[1]{\ensuremath{\left|#1\right\rangle}}
\definecolor{blue}{rgb}{0,0,1}
\definecolor{red}{rgb}{1,0,0}
\definecolor{green}{rgb}{0,1,0}

\usepackage[unicode=true,pdfusetitle,
 bookmarks=true,bookmarksnumbered=false,bookmarksopen=false,
 breaklinks=false,pdfborder={0 0 0},pdfborderstyle={},backref=false,colorlinks=true]
 {hyperref}
\hypersetup{
 colorlinks,linkcolor=red,citecolor=blue}
\makeatother

\begin{document}

\title{Tunable coupler for realizing a controlled-phase gate with dynamically decoupled regime in a superconducting circuit}

\author{X. Li}
\thanks{These two authors contributed equally to this work.}
\affiliation{Center for Quantum Information, Institute for Interdisciplinary Information Sciences, Tsinghua University, Beijing 100084, China}

\author{T. Cai}
\thanks{These two authors contributed equally to this work.}
\affiliation{Center for Quantum Information, Institute for Interdisciplinary Information Sciences, Tsinghua University, Beijing 100084, China}

\author{H. Yan}
\affiliation{Center for Quantum Information, Institute for Interdisciplinary Information Sciences, Tsinghua University, Beijing 100084, China}

\author{Z. Wang}
\affiliation{Center for Quantum Information, Institute for Interdisciplinary Information Sciences, Tsinghua University, Beijing 100084, China}

\author{X. Pan}
\affiliation{Center for Quantum Information, Institute for Interdisciplinary Information Sciences, Tsinghua University, Beijing 100084, China}

\author{Y. Ma}
\affiliation{Center for Quantum Information, Institute for Interdisciplinary Information Sciences, Tsinghua University, Beijing 100084, China}

\author{W. Cai}
\affiliation{Center for Quantum Information, Institute for Interdisciplinary Information Sciences, Tsinghua University, Beijing 100084, China}

\author{J. Han}
\affiliation{Center for Quantum Information, Institute for Interdisciplinary Information Sciences, Tsinghua University, Beijing 100084, China}

\author{Z. Hua}
\affiliation{Center for Quantum Information, Institute for Interdisciplinary Information Sciences, Tsinghua University, Beijing 100084, China}

\author{X. Han}
\affiliation{Center for Quantum Information, Institute for Interdisciplinary Information Sciences, Tsinghua University, Beijing 100084, China}

\author{Y. Wu}
\affiliation{Center for Quantum Information, Institute for Interdisciplinary Information Sciences, Tsinghua University, Beijing 100084, China}

\author{H. Zhang}
\affiliation{Center for Quantum Information, Institute for Interdisciplinary Information Sciences, Tsinghua University, Beijing 100084, China}

\author{H. Wang}
\affiliation{Center for Quantum Information, Institute for Interdisciplinary Information Sciences, Tsinghua University, Beijing 100084, China}

\author{Yipu Song}\email{ypsong@mail.tsinghua.edu.cn}
\affiliation{Center for Quantum Information, Institute for Interdisciplinary Information Sciences, Tsinghua University, Beijing 100084, China}

\author{Luming Duan}\email{lmduan@tsinghua.edu.cn}
\affiliation{Center for Quantum Information, Institute for Interdisciplinary Information Sciences, Tsinghua University, Beijing 100084, China}

\author{Luyan Sun}\email{luyansun@tsinghua.edu.cn}
\affiliation{Center for Quantum Information, Institute for Interdisciplinary Information Sciences, Tsinghua University, Beijing 100084, China}

\begin{abstract}

Controllable interaction between superconducting qubits is desirable for large-scale quantum computation and simulation. Here, based on a theoretical proposal by Yan $et~al$. [Phys. Rev. Appl. \textbf{10}, 054061 (2018)] we experimentally demonstrate a simply-designed and flux-controlled tunable coupler with a continuous tunability by adjusting the coupler frequency, which can completely turn off adjacent superconducting qubit coupling. Utilizing the tunable interaction between two qubits via the coupler, we implement a different type of controlled-phase (CZ) gate with `dynamically decoupled regime', which allows the qubit-qubit coupling to be only `on' at the usual operating point while dynamically `off' during the tuning process of one qubit frequency into and out of the operating point. This scheme not only efficiently suppresses the leakage out of the computational subspace, but also allows for the acquired two-qubit phase being geometric at the operating point. We achieve an average CZ gate fidelity of 98.3$\pm0.6$\%, which is dominantly limited by qubit decoherence. The demonstrated tunable coupler provides a desirable tool to suppress adjacent qubit coupling and is suitable for large-scale quantum computation and simulation.

\end{abstract}

\maketitle

\section{Introduction}
As enormous progress has been made towards more complex networks of qubits~\cite{Otterbach2017,Kandala2017,Google2019,Song2019,Yan2019,Ma2019,Kollar2019,andersen2019repeated}, superconducting quantum circuits have become a promising implementation for quantum simulation~\cite{Buluta2009,Houck2012,Georgescu2014} and fault-tolerant quantum computation~\cite{Devoret2013,Campbell2017}. Building large circuits requires long coherent times of the qubit, strong interqubit interaction for fast and high-fidelity two-qubit gates, and small to zero coupling between qubits when no interaction is needed. For typical planar circuits with transmon or transmon-type qubits connected through fixed capacitors or quantum buses, strong interaction and variable coupling can be achieved by dynamically adjusting the frequencies of the tunable qubits~\cite{Majer2007,DiCarlo2009,Kelly2015,Song2019} or by applying external microwave drives~\cite{Chow2011,Sheldon2016}. Parametric modulation of the frequency of the qubit or the bus has also been used to achieve tunable coupling between qubits~\cite{Zhou2009,Strand2013,Liu2014,Xue2015,Wu2016,Caldwell2018,Reagor2018,Li2018,Chen2018Nonadiabatic,Cai2019,Xu2019}. However, these approaches could not fully turn off the interaction, and thus parasitic ZZ crosstalk coupling is always present, resulting in a frequency shift of one qubit depending on the state of another. This unwanted qubit interaction could be a limited source to degrade the single-qubit gate performance and to accumulate the entanglement phase error. In addition, because of the requirement of the qubit frequency tunability or the relatively small qubit-qubit detunings, these approaches also suffer from the frequency-crowding problem.

Inserting an extra circuit element, a tunable coupler, can offer another degree of freedom, and thus is efficient for helping mitigate the above problems of unwanted interactions and frequency crowding, while achieving a controllable qubit interaction without introducing other nonidealities that limit the gate performance. A variety of tunable couplers based on a rf superconducting quantum interference device, a tunable bus, or a tunable inductor have previously been designed and demonstrated experimentally~\cite{Liu2006,Niskanen2007,Ploeg2007,Harris2007,Allman2010,Bialczak2011,Chen2014,Mckay2016,Lu2017,Kounalakis2018tuneable,Chen2018,Neill2018,Mundada2019}. Tunable couplers are thus desirable for scalable architectures for quantum computation and simulation applications.

In this work, following the theoretical proposal in Ref.~\onlinecite{Yan2018}, we experimentally demonstrate a simply designed tunable coupler capacitively coupled to two computational Xmon qubits in a superconducting circuit~\cite{You2011,Devoret2013,Gu2017,Krantz2019}. This tunable coupler is based on only one extra Xmon qubit and is therefore easy to scale up. By adjusting the coupler frequency, the qubit-qubit coupling strength can be tuned through a combination of different coupling paths such that a continuous tunability from positive to negative values can be realized. Consequently, unwanted qubit interactions, such as parasitic ZZ crosstalk, can be completely turned off as wished. Utilizing the tunable interaction between the two qubits via the coupler, we realize the entangling gates of iswap and $\sqrt{\mathrm{iswap}}$ with a fidelity of 96.8\% and 95.0\%, respectively. In addition, we implement a different type of controlled-phase (CZ) gate by using a `dynamically decoupled regime' (DDR) technique, which allows the qubit-qubit coupling to be only `on' at the usual operating point (where the two qubit states $\ket{11}$ and $\ket{20}$ are resonant), while dynamically `off' during the tuning process of one qubit into and out of the operating point by simultaneously tuning the coupler frequency. Compared to the CZ gate implemented with a rectangular pulse or a fast adiabatic pulse~\cite{martinis2014fast}, this scheme not only efficiently suppresses the leakage out of the computational subspace, but also allows for a geometric $\pi$-phase accumulation on $\ket{11}$ state, which is potentially more robust~\cite{Zhu2005}. We achieve an average CZ gate fidelity of 98.3$\pm0.6$\%, which is dominantly limited by qubit decoherence. 

Besides, the demonstrated tunable coupler provides a straightforward way to suppress adjacent qubit coupling without degrading the qubit coherence. Theoretically, this tunable coupler scheme can offer a wide coupling tunability from several MHz in the positive regime to several tens MHz in the negative regime, thus allowing for fast two-qubit gates. The simple design of adding only one extra Xmon qubit as the coupler is also particularly compatible and desirable for large-scale superconducting architectures.

\begin{figure}[t]
\includegraphics{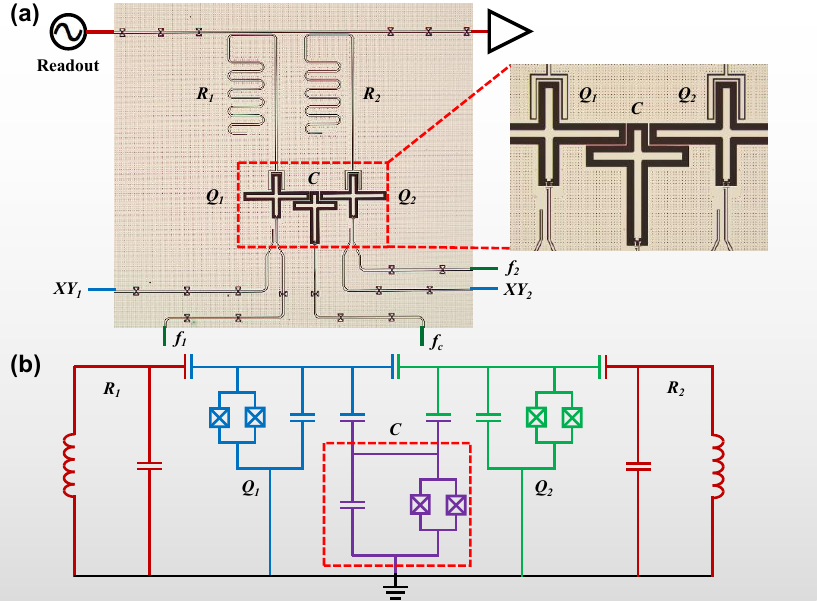}
\caption{(a) Optical micrograph of three Xmon qubits with the middle one $C$ serving as a tunable coupler for the two computational qubits ($Q_1$ and $Q_2$). Each computational qubit has independent XY and Z control, and is coupled to a separate $\lambda/4$ resonator for simultaneous and individual readout. The coupler also has an individual flux-bias line for a frequency tunability. The combination of direct capacitive coupling and indirect tunable coupling via the coupler constitutes the total coupling between the two computational qubits. (b) Schematic electrical circuit of the device.}
\label{fig:Fig1}
\end{figure}

\begin{figure}[ht]
\includegraphics{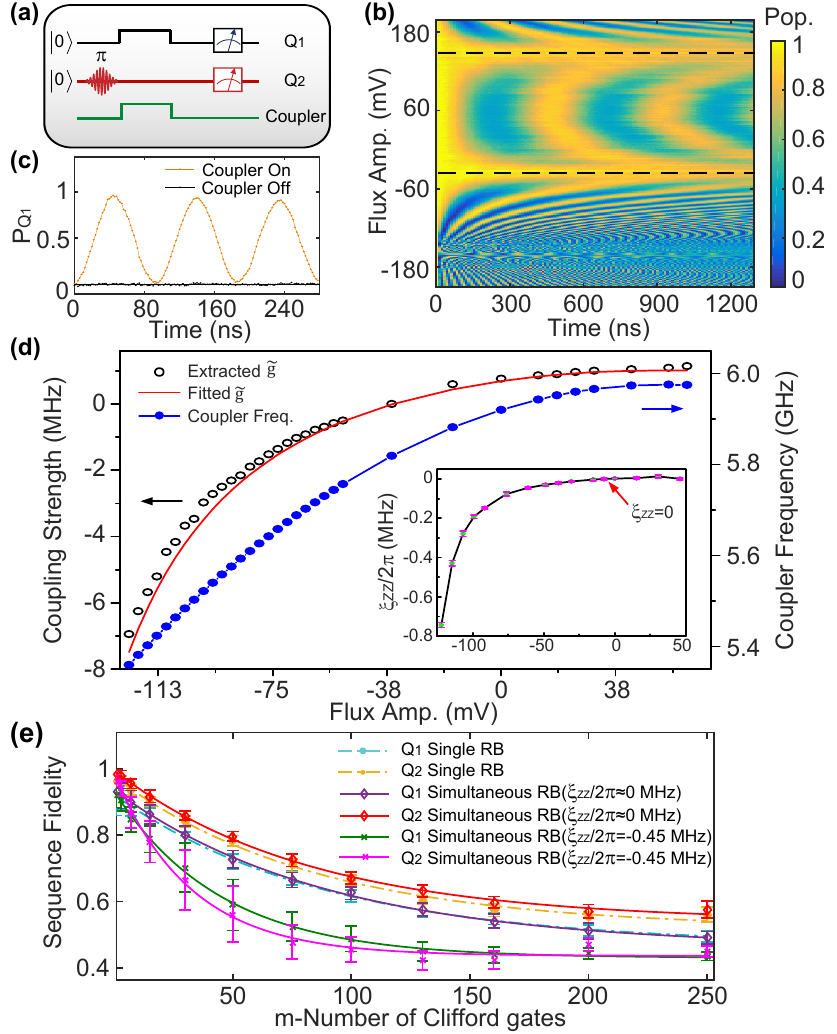}
\caption{(a) Pulse sequence to characterize the tunability of the coupler. The two qubits are initialized in the ground state at their sweet spots with a detuning of 35~MHz. The coupler is originally set at $\omega_{c}/2\pi=5.905$~GHz where the coupling between the two qubits is nearly off. A $\pi$ pulse is first to excite $Q_2$, followed by two simultaneous fast flux pulses: $f_1$ brings $Q_1$ into resonance with $Q_2$; $f_c$ on the coupler to turn on the coupling. After the two qubits interact and evolve for time $t$, $Q_1$ and the coupler are pulsed back to the original points for measurements of qubit populations. (b) Population of $Q_2$ as a function of the amplitude of $f_c$ and $t$ clearly reveals the tunability of the coupling strength. The two dark dashed lines indicate the situation where the coupling is off. (c) Population of $Q_1$ as a function of time with the coupling on (orange dots) or off (black dots). (d) The effective qubit-qubit coupling strength $\tilde{g}/2\pi$ (black circles) extracted by fitting the oscillation of the qubit excitation in (b) as a function of the flux-bias amplitude on the coupler. The red line is a fit to the extracted $\tilde{g}$ according to Eq.~(\ref{Eq3}). The coupler frequency (blue dots) can be measured independently by probing the dispersive shift of the qubit frequency when pulsing the coupler into the excited state. Inset: the ZZ crosstalk coupling $\xi_{ZZ}/2\pi$ measured in a Ramsey-type experiment when the two qubits are detuned and at their sweet spots. The red arrow indicates where the coupling is off. (e) Individual and simultaneous RB for $Q_1$ and $Q_2$ with $\xi_{ZZ}/2\pi\approx0$ and $-0.45$~MHz, respectively.}
\label{fig:Fig2}
\end{figure}

\section{Results}
\subsection{Experimental System and Theory}
Our experimental device consists of three flux-tunable Xmon qubits ($Q_1, C, Q_2$)~\cite{Barends2013,Barends2014,Kelly2015} with the middle one $C$ serving as the tunable coupler (henceforth referred to as the `coupler'), as shown in Fig.~\ref{fig:Fig1}(a). Figure~\ref{fig:Fig1}(b) is the schematic of the device. The maximum frequencies of the two qubits and the coupler are $\omega^{\mathrm{max}}_{1}/2\pi=4.961$~GHz, $\omega^{\mathrm{max}}_{2}/2\pi=4.926$~GHz, and $\omega^{\mathrm{max}}_{c}/2\pi=5.977$~GHz. Details of the experimental apparatus and device parameters are presented in  Appendix C. We first briefly discuss the operating principle of the tunable coupler following Ref.~\onlinecite{Yan2018}. The system can be described by the Hamiltonian:
\begin{equation}\label{Eq1}
\begin{split}
H/\hbar&=\sum_{i=1,2}\frac{1}{2}{\omega_i}\sigma^z_i+\frac{1}{2}{\omega_c}\sigma^z_c+\sum_{i=1,2}g_i(\sigma^+_i\sigma^-_c+\sigma^+_c\sigma^-_i)\\
&+g_{12}(\sigma^+_1\sigma^-_2+\sigma^+_2\sigma^-_1),
\end{split}
\end{equation}
where $\omega_\alpha (\alpha=1,2,c)$ are the frequencies of $Q_1$, $Q_2$, and the coupler respectively, $\sigma_\alpha^{z,\pm}$ are the corresponding Pauli Z, raising and lowering operators, $g_i$ (i = 1, 2) is the coupling strength between each qubit and the coupler, $g_{12}$ is the direct capacitive coupling strength between the two qubits.

In the strong dispersive regime ($g_i\ll|\Delta_i|$, where $\Delta_i=\omega_i-\omega_c)$ and assuming that the coupler mode remains in its ground state, an effective two-qubit Hamiltonian with the variable coupler decoupled can be derived by making the unitary transformation $U=\mathrm{exp}\{\sum_{i=1,2} g_i/\Delta_i (\sigma^{+}_i \sigma^{-}_c-\sigma^{-}_i\sigma^{+}_c)$\}~\cite{Blais2007,Bravyi2011} and keeping to second order in $g_i/\Delta_i$:
\begin{equation}\label{Eq2}
UHU^{\dagger}/\hbar=\sum_{i=1,2}\frac{1}{2}{\tilde{\omega}_i}\sigma^z_i+\tilde{g}(\sigma^+_1\sigma^-_2+\sigma^-_1\sigma^+_2),
\end{equation}
where ${\tilde{\omega}_i}= {\omega}_i+g_i^2/\Delta_i$ is the dressed frequency and
\begin{equation}\label{Eq3}
\tilde{g} =g_{12}+(g_1 g_2)/\Delta
\end{equation}
is the effective coupling strength with $1/\Delta=(1/\Delta_1 +1/\Delta_2)/2$. The interaction between the two qubits thus consists of the direct capacitive coupling and the indirect virtual exchange coupling via the coupler. If $\Delta_i<0$ (the coupler's frequency is above both qubits' frequencies), the virtual exchange interaction term $(g_1 g_2)/\Delta<0$. Therefore, the effective coupling $\tilde{g}$ can be tuned from negative to positive monotonically with increasing the coupler frequency. Since this coupling tunability is continuous, a critical value $\omega_c^{\mathrm{off}}$ can always be reached to turn off the qubit-qubit coupling $\tilde{g}(\omega_c^{\mathrm{off}})=0$. When the two qubits are detuned, the ZZ crosstalk coupling $\xi_{ZZ}$ can also be turned off (see below and  Appendix A).

\subsection{Suppressing Parasitic ZZ Crosstalk}
We now demonstrate the tunability of the qubit-qubit coupling strength $\tilde{g}$ controlled by the coupler's frequency. The experimental pulse sequence is illustrated in Fig.~\ref{fig:Fig2}(a). Coherent excitation oscillation between $Q_1$ and $Q_2$ as a function of the flux amplitude of $f_c$ on the coupler and time $t$ is shown in Fig.~\ref{fig:Fig2}(b), and clearly reveals the change of the coupling strength depending on the coupler frequency. Remarkably, two flux biases of the coupler, at which the qubit-qubit interaction is completely turned off, are observed and marked by two dark dashed lines.

The extracted $\tilde{g}$ indeed varies continuously from positive to negative values and is in good agreement with theoretical calculations (red curve), as shown in Fig.~\ref{fig:Fig2}(d). The small discrepancy at large negative coupling regime owes to the deviation of the qubit-coupler coupling from the strong dispersive condition. Given the extracted coupling strength and the frequency detuning of each qubit from the coupler, we can estimate the direct capacitive coupling strength $g_{12} \approx 6.74$~MHz using Eq.~(\ref{Eq3}). We note that a large negative interaction can be reached with the decrease of the coupler frequency approaching the qubit frequency. However, when the coupler frequency is further reduced to be close to the qubit frequency, the virtual excitation approximation of the coupler becomes invalid.

\begin{figure}
\includegraphics{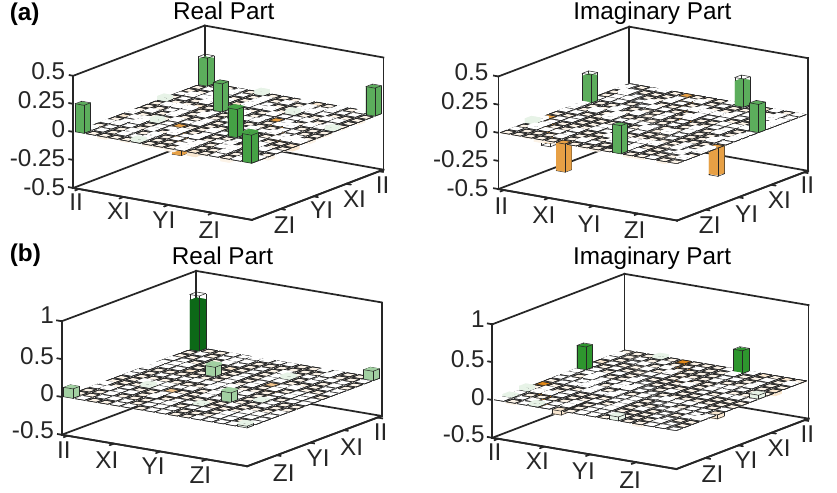}
\caption{(a) (b) Bar charts of the measured $\chi_{\mathrm{exp}}$ from a QPT for iswap and $\rm \sqrt{iswap}$ with a gate fidelity of 96.8\% and 95.0\%, respectively. The solid black outlines are for the ideal gate.}
\label{fig:Fig3}
\end{figure}

We extract $\xi_{ZZ}$ when the two qubits are detuned at their sweet spots using a Ramsey-type experiment, which involves probing the frequency of $Q_2$ with $Q_1$ in either its ground or excited state~\cite{reed2013entanglement,chow2010quantum}. The measured $\xi_{ZZ}$ also depends on the coupler frequency and is shown in the inset of Fig.~\ref{fig:Fig2}(d). At the critical coupler frequency $\omega_c^{\mathrm{off}}\approx 5.905$~GHz, indicated by the red arrow, the measured $\xi_{ZZ}/2\pi\approx 1$~kHz and is limited by the current detection scheme. We utilize simultaneous randomized benchmarking (RB) to verify the isolation of two qubits at this configuration. The simultaneous RB gate fidelities of 99.45\% and 99.40\% are nearly the same as the individual RB gate fidelities of 99.44\% and 99.41\% for $Q_1$ and $Q_2$ respectively, as shown in Fig.~\ref{fig:Fig2}(e). For comparison, when the two qubits are biased in the same configuration as above but with $\xi_{ZZ}/2\pi$=$-0.45$~MHz, the simultaneous RB gate fidelities on both qubits are lowered by about 0.54\% ($Q_1$) and 0.93\% ($Q_2$). This contrast illustrates the importance of the tunable coupler for precise qubit control.

\subsection{Implementing CZ Gate with Dynamically Decoupled Regime}
\begin{figure}
\includegraphics{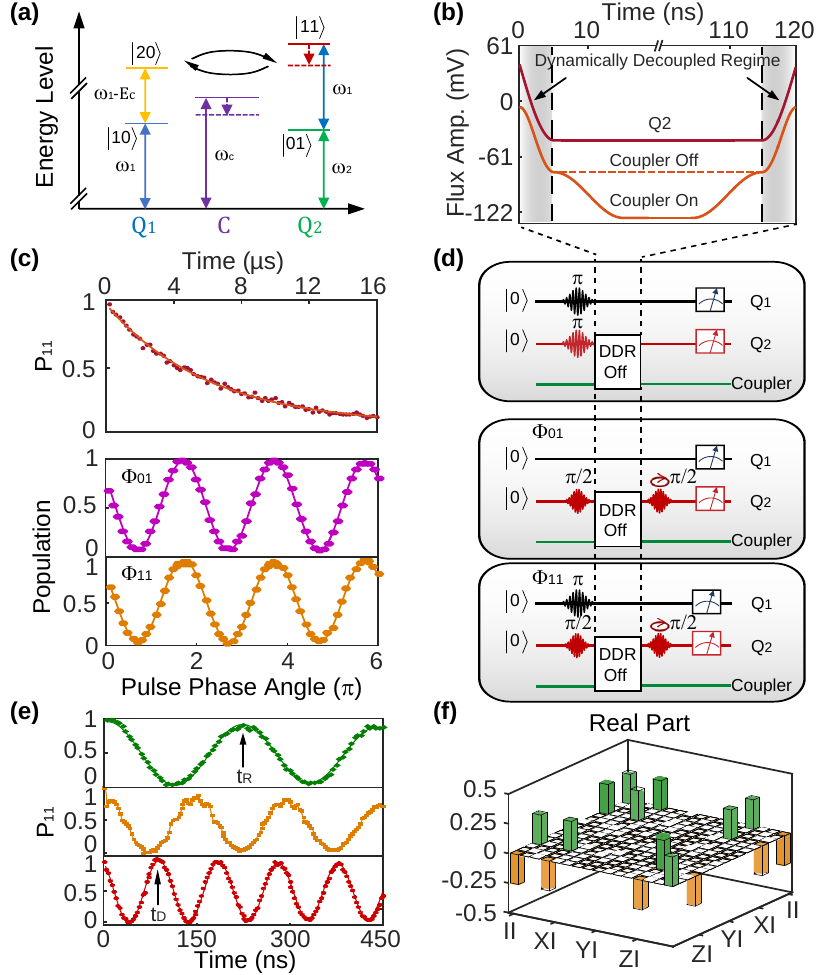}
\caption{(a) Schematic to realize the CZ gate using the usual $\ket{11}$ and $\ket{20}$ resonance. (b) Flux sequence to realize the CZ gate with DDR. The $Q_1$ frequency is unchanged during the process. A gradual flux pulse (for a cosine-type frequency adjustment) on $Q_2$ is to tune its frequency to the operating point, while a flux pulse on the coupler with an appropriate pulse shape to dynamically keep the coupling `off' (the shaded regions). The coupling is then turned on by applying another gradual flux pulse on the coupler (for a cosine-type coupler frequency adjustment) for proper two-qubit interaction time to acquire the required $\pi$-phase shift on $\ket{11}$. (c), (d) Confirmation of the `off' state of the coupling in the DDR and the corresponding measurement sequences. Top: the population of $\ket{11}$ state shows purely exponential decay without any oscillation. Middle: phase accumulation of $\ket{01}$ with respect to $\ket{00}$. Bottom: phase accumulation of $\ket{11}$ with respect to $\ket{10}$. The difference between the lower two panels shows zero phase accumulation of $\ket{11}$, confirming zero coupling. (e) Population of $\ket{11}$ in three different schemes to realize the CZ gate. Upper: rectangular pulses to have a positive coupling strength. $t_R=222$~ns is required for realizing the standard two-qubit CZ gate. Middle: rectangular pulses to have a negative coupling regime. Obvious high-frequency oscillations occur and indicate the leakage out of $\ket{11}$. Lower: DDR pulses to have a large negative coupling but with a smooth oscillation. A CZ gate can be accomplished with a gate time $t_D$=119~ns. (f) Bar chart of the measured real part of $\chi_{\mathrm{exp}}$ from a QPT shows a gate fidelity of 98.3$\pm0.6$\%. The solid black outlines are for the ideal gate.}
\label{fig:Fig4}
\end{figure}

With the tunability of the coupling, we now move to the implementation of two-qubit entangling gates. The iswap and $\rm \sqrt{iswap}$ gates are quite natural based on the Hamiltonian of Eq.~(\ref{Eq2})~\cite{Mckay2016}. Their measured $\chi_{\mathrm{exp}}$ from quantum-process tomography (QPT), which can give full information about the gate process~\cite{korotkov2013}, are presented in Fig.~\ref{fig:Fig3} with a gate fidelity of 96.8\% and 95.0\% respectively. Here we focus on the controlled-phase (CZ) gate. The CZ gate is implemented by using the usual avoided crossing of the noncomputational state $\ket{20}$ with the $\ket{11}$ state [Fig.~\ref{fig:Fig4}(a)], which is only accessible to $\ket{11}$ and thus provides the conditional nature of the gate to flip the phase if and only if both qubits are excited~\cite{Strauch2003,DiCarlo2009,Reed2012,Li2019nonadiabaticCZ}.

The fully controllable interaction of the coupler allows for turning the coupler `on' or `off' as wished. The ideal scheme of implementing a CZ gate with a coupler is to have both qubits initialized in the operating point where $\ket{11}$ and $\ket{20}$ have the same energy with the coupling `off', and then slowly turn on the coupler for proper time to implement the gate. This method can avoid adjusting the qubit frequency, and thus reduce the leakage error. However, the unwanted crosstalk of the XY control lines in our device could degrade single-qubit gate performance because of the zero detuning between $\omega_{12}$ of $Q_1$ and $\omega_{01}$ of $Q_2$. Therefore, the qubits are initially detuned and at their sweet spots with the coupling `off'.

The simplest case is to use rectangular fluxes to simultaneously pulse the qubits to the avoided-crossing point and turn on the coupling. Both our measurement and simulation reveal that the leakage out of $\ket{11}$ can be effectively suppressed in the positive coupling regime rather than the negative one. This is evidenced by the observation of high-frequency oscillations [the middle panel in Fig.~\ref{fig:Fig4}(e)], which cause unwanted leakage error. However, due to the weak positive coupling strength, a relatively long gate time ($t_R=222$~ns) is required to accomplish the CZ gate. The average CZ gate fidelity is 95.5\% from the QPT measurement. 

To improve the gate fidelity, we use a `dynamically decoupled regime' (DDR) scheme to implement the CZ gate. Note that DDR defined here differs in explication from the term `dynamical decoupling' used in coherent control pulse methods to suppress the dephasing error~\cite{Viola1998}. This scheme can not only efficiently suppress the leakage out of the computational subspace but also allow us to perform the gate in the negative coupling regime with larger interaction strength for a shorter gate time. The DDR scheme can be understood by mapping the interaction between the states $\ket{11}$ and $\ket{20}$ onto a Hamiltonian of a qubit $H=H_x\sigma_x'+H_z\sigma_z'$~\cite{martinis2014fast}, where $H_x$ is the coupling energy between the states $\ket{11}$ and $\ket{20}$, $H_z$ is the frequency detuning of $\ket{11}$ from the resonance point of the avoided crossing, and $\sigma_x'$ and $\sigma_z'$ are the corresponding Pauli operators which are distinguished from those on the computational qubit. For the typical direct coupling scheme without a coupler where the coupling energy $\sigma_x'$ term is fixed, CZ gate is realized by adjusting only the $\sigma_z'$ term, such as the one implemented with a rectangular pulse, fast adiabatic pulse or nonadiabatic pulse~\cite{martinis2014fast,Li2019nonadiabaticCZ}. While for the DDR scheme with a coupler, both $\sigma_x'$ term and $\sigma_z'$ term can be varied as wished.

The experimental flux sequence is shown in Fig.~\ref{fig:Fig4}(b). We use a gradual flux pulse on $Q_2$ to tune its frequency to the operating point, while dynamically keeping the coupling `off'. This DDR is achieved by applying a flux pulse on the coupler with an appropriate pulse shape calculated by Eq.~(\ref{Eq3}) and is further optimized to assure the zero coupling during the whole qubit frequency-changing process [the shaded regions where $Hx=0, Hz=Hz(t)$]. After that, we slowly turn on the coupling by applying another gradual flux pulse on the coupler and wait for proper time for the two-qubit interaction to acquire the $\pi$-phase shift on $\ket{11}$ [$Hx=Hx(t), Hz=0$]. The `off' state of the coupling in the DDR is confirmed by measuring the population change and phase accumulation of $\ket{11}$ state with the same flux pulses as in Fig.~\ref{fig:Fig4}(b) (the DDR and the dashed line in the middle). The experimental results and sequences are shown in Figs.~\ref{fig:Fig4}(c) and \ref{fig:Fig4}(d), respectively.

We choose an appropriate negative coupling strength to perform the CZ gate, balancing the gate time and qubit-coupler leakage induced by the non-negligible excitation of the coupler. The larger interaction strength reduces the CZ gate time to $t_D=119$~ns [bottom panel in Fig.~\ref{fig:Fig4}(e)] and the gate fidelity is improved with an average QPT fidelity of 98.3$\pm0.6$\%. The measured $\chi_{\mathrm{exp}}$ is shown in Fig.~\ref{fig:Fig4}(f). Moreover, the observed two-qubit $\pi$-phase accumulation is geometric, which is potentially noise resilient to frequency fluctuation during the gate operation. The coupler frequency can be further lowered for a higher coupling strength such that the CZ gate time can be reduced to $t_{D}=68$~ns. However, due to a larger leakage between the qubit and the coupler, the gate fidelity is slightly lower. It is worth mentioning that we also consider a synchronization optimization strategy to mitigate leakages from not only the qubit-qubit $\ket{01}$ and $\ket{10}$ swap channel but also the qubit-coupler real-energy exchange channel~\cite{Barends2019}.

\section{Discussion}
Because the tunable coupler provides an extra degree of freedom and can fully suppress the qubit coupling, the CZ gate with the DDR scheme should be insensitive to qubit parameters. Based on a numerical simulation (see Appendix E), in the absence of qubit decoherence, the QPT fidelity of the CZ gate with the DDR scheme can be above $99.99\%$ (with `mesolve' in QuTip~\cite{johansson2012,johansson2013}). For comparison, we also numerically simulate the CZ gate implemented for the direct coupling case with a fast adiabatic pulse~\cite{martinis2014fast}. We choose the same qubit idle frequencies and qubit-qubit coupling strength to ensure the same gate time ($\sim 120$ ns) as the DDR scheme, and include the first three coefficients of Fourier basis functions in the pulse optimization with the `fmin' function in PYTHON. The resulted maximum CZ gate fidelity $F \approx 99.60\%$, with an error significantly larger than that from the DDR scheme, because the fast adiabatic pulse is more sensitive to qubit parameters and pulse optimization. We note that more Fourier coefficients in the fast adiabatic pulse can be included to improve the CZ gate fidelity, but at the expense of more optimization parameters and harder optimization. A high-fidelity, fast adiabatic CZ gate would be even more challenging because of the unwanted crosstalk when the qubit system becomes larger. By contrast, the CZ gate with the DDR scheme should maintain its robustness and convenience, and therefore would be particularly suitable for large-scale superconducting circuits.
 
In current experiment, the measured CZ gate fidelity is dominantly limited by the qubit decoherence, while the coupler decoherence has little effect on the gate fidelity since the coupler remains in the ground state. Optimization of coupler design for better parameters would be helpful to achieve a higher coupling strength while decreasing the leakage error and consequently get a shorter CZ gate time with a improved gate performance. Besides, other pulse shapes to turn on the coupling may further suppress the leakage and need future exploration. In addition, improving fabrication technology to minimize crosstalk between XY lines can offer possibility of realizing a more ideal CZ gate~\cite{dunsworth2018method}. Finally, by adopting the Nelder-Mead optimization protocol~\cite{Kelly2014}, we could further improve the CZ gate fidelity to a higher level.

In summary, we experimentally realize a simple prototype of a flux-controlled tunable coupler. The competition between the positive direct and negative indirect coupling allows for a continuous tunability and for switching off the coupling completely. With this coupler, we implement two-qubit entangling iswap, $\rm \sqrt{iswap}$, and CZ gates. In particular, the CZ gate is realized with fully dynamical control over the qubit-qubit coupling: the coupling is only on at the operating point to acquire a geometric two-qubit phase, while being off during the tuning process of the qubit frequency. We achieve an average CZ gate fidelity of 98.3$\pm0.6$\%, characterized via QPT and dominantly limited by system decoherence. The demonstrated tunable coupler therefore provides a desirable tool to suppress adjacent qubit coupling and is suitable for large-scale quantum computation and simulation~\cite{Buluta2009,Houck2012,Georgescu2014,Kyriienko2018}.

\section*{Acknowledgments}
We thank Chengyao Li for the technical support, and Changling Zou and Zhengyuan Xue for valuable discussions. This work is supported by National Key Research and Development Program of China under Grant No. 2016YFA0301902 and 2017YFA0304303, and National Natural Science Foundation of China under Grant No.11874235 and No.11925404.

\appendix

\section{Calculation of the effective qubit coupling $\widetilde{g}$ and ZZ crosstalk coupling $\xi_{ZZ}$}
To study the tunable coupling strength between $Q_1$ and $Q_2$, we consider the system Hamiltonian,

\begin{figure}[b]
\includegraphics{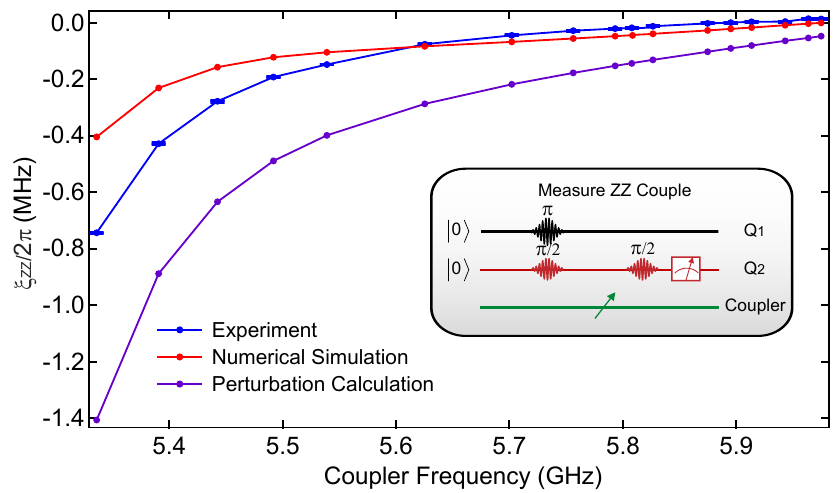}
\caption{ZZ crosstalk coupling $\xi_{ZZ}$ versus coupler frequency when the two qubits are detuned and at their sweet spots as in the inset of Fig.~\ref{fig:Fig2} of the main text. Inset: the experimental sequence for measuring ZZ crosstalk coupling $\xi_{ZZ}$ at a specific coupler flux bias.}
\label{fig:ZZcouplingStrength}
\end{figure}

\begin{equation}\label{Eq_System_Hamiltonian}
\begin{split}
\widehat{H}=\widehat{H}_0&+\widehat{V}\\
\widehat{H}_0/\hbar=&\sum_{i=1,2,c} \omega_ia_i^+a_i+\frac{\eta_i}{2}a_i^+a_i^+a_ia_i\\
\widehat{V}/\hbar=&\sum_{j=1,2} g_{jc}(a_j^+a_c+a_ja_c^+)+g_{12}(a_1^+a_2+a_1a_2^+)
\end{split}
\end{equation}
where $\omega_i$ and $\eta_i$ ($i=1,2,c$) are the frequencies and anharmonicities of $Q_1$, $Q_2$, and the coupler respectively, $a_i^+$ and $a_i$ are the corresponding raising and lowering operators, $g_{jc}$ ($j$ = 1, 2) is the coupling strength between each qubit and the coupler, and $g_{12}$ is the direct capacitive coupling strength between the two qubits.

When we only consider the ground and the first excited states of the Xmon qubits, based on the Schrieffer-Wolff transformation~\cite{Blais2007,Bravyi2011}, the effective coupling $\widetilde{g}$ between $Q_1$ and $Q_2$ would be~\cite{Yan2018},
\begin{equation}\label{Eq_XXcoupling}
\begin{split}
\widetilde{g}=\frac{g_{2c}g_{1c}}{2\Delta_{2c}}+\frac{g_{2c}g_{1c}}{2\Delta_{1c}}+g_{12},
\end{split}
\end{equation}
where $\Delta_{ic} \equiv \omega_i-\omega_c$ is the frequency detuning between the qubit and the coupler.

To calculate the parasitic ZZ crosstalk coupling $\xi_{ZZ}$ between $Q_1$ and $Q_2$ when they are detuned, however, it is not enough to only keep up to the second order in the Schrieffer-Wolff transformation. Here we use the perturbation approach~\cite{zhu2013circuit,krishnan1978approximate} to derive the parasitic ZZ crosstalk coupling to second, third, and fourth order of the Hamiltonian (\ref{Eq_System_Hamiltonian}). We define $\xi_{ZZ}=\omega_{11}-\omega_{01}-\omega_{10}$ between $Q_1$ and $Q_2$. The corresponding perturbation terms are:
\begin{equation}\label{Eq9}
\xi_{ZZ}^{(2)}=\frac{2(g_{12})^2(\eta_1+\eta_2)}{(\Delta_{12}+\eta_1)(\Delta_{12}-\eta_2)},
\end{equation}
\begin{equation}\label{Eq10}
\begin{split}
\xi_{ZZ}^{(3)}=2g_{12}&g_{1c}g_{2c}[\frac{1}{\Delta_{2c}}(\frac{2}{\Delta_{21}-\eta_1}-\frac{1}{\Delta_{21}})\\
+&\frac{1}{\Delta_{1c}}(\frac{2}{\Delta_{12}-\eta_2}-\frac{1}{\Delta_{12}})],
\end{split}
\end{equation}
\begin{equation}\label{Eq11}
\begin{split}
\xi_{ZZ}^{(4)}=&\frac{2(g_{1c})^2(g_{2c})^2}{\Delta_{1c}+\Delta_{2c}-\eta_c}(\frac{1}{\Delta_{1c}}+\frac{1}{\Delta_{2c}})^2\\
&+\frac{(g_{1c})^2(g_{2c})^2}{(\Delta_{1c})^2}(\frac{2}{\Delta_{12}-\eta_2}-\frac{1}{\Delta_{12}}-\frac{1}{\Delta_{2c}})\\
&+\frac{(g_{1c})^2(g_{2c})^2}{(\Delta_{2c})^2}(\frac{2}{\Delta_{21}-\eta_1}-\frac{1}{\Delta_{21}}-\frac{1}{\Delta_{1c}}),
\end{split}
\end{equation}
where $\Delta_{ij} \equiv \omega_i-\omega_j$ ($i,j=1,2,c; i\neq j$). Because $g_{12}\ll g_{ic}$, we omit the $g_{12}$ term in the derivation of the fourth-order perturbation. Finally, $\xi_{ZZ}=\xi_{ZZ}^{(2)}+\xi_{ZZ}^{(3)}+\xi_{ZZ}^{(4)}$ and the perturbation calculation is compared with the experiment and the numerical simulation based on QuTip~\cite{johansson2012,johansson2013}, as shown in Fig.~\ref{fig:ZZcouplingStrength}. The experiment agrees fairly well with the numerical simulation. The deviation between the experiment and the perturbation calculation may be due to the higher-order terms. When the detuning between the coupler and the qubits decreases, we can see a larger deviation, indicating the higher-order perturbation terms become more important. The experiment, simulation, and perturbation calculation all reveal that $\xi_{ZZ}$ can be tuned from negative to positive continuously, and thus we can always find a critical point to fully turn off $\xi_{ZZ}$ between $Q_1$ and $Q_2$.

\section{Quantum Process Tomography}
The two-qubit quantum gates are characterized with quantum-process tomography (QPT). Generally, the qubits are initialized to the following 16 states \{$\ket{g}$, $\ket{e}$, $(\ket{g}+\ket{e})/\sqrt{2}$, $(\ket{g}-i\ket{e})/\sqrt{2}$\}$^{\otimes 2}$ with the proper single-qubit rotations. After the gate that needs to be characterized, the corresponding final two-qubit state is reconstructed from state tomography measurements with 16 prerotations \{$I, X/2, Y/2, X$\}$^{\otimes 2}$, where $I$ is the identity operator, $X, Y, X/2$, and  $Y/2$ are single-qubit $\pi$ and $\pi$/2 rotations around X and Y axes respectively. With the 16 initial states $\rho_i$, the experimental process matrix $\chi_{\mathrm{exp}}$ can be extracted from the 16 corresponding final states $\rho_f$ through $\rho_f=\sum_{m,n}{\chi_{mn} E_m\rho_i E_n^{\dagger}}$~\cite{nielsen2002quantum}, where the basis operators $E_m$ and $E_n$ are chosen from the set \{$I$, $\sigma_x$, $-i\sigma_y$, $\sigma_z$\}$^{\otimes 2}$.

However, the real experiment is not perfect. For example, the generated $\rho_i$ are not ideal due to the initial state preparation errors; the final state tomography could also be nonideal due to the readout errors. To solve this problem, we first use state tomography to characterize the preparation of the initial states. The measured initial states $\rho_i^{\mathrm{meas}}$ are then used to extract $\chi_{\mathrm{exp}}$ through $\rho_f=\sum_{m,n}{\chi_{mn} E_m\rho_i^{\mathrm{meas}} E_n^{\dagger}}$. The gate-process fidelity is finally calculated through $F = Tr (\chi_{\mathrm{exp}}\chi_{\mathrm{ideal}})$, where $\chi_{\mathrm{ideal}}$ is the ideal process matrix for the corresponding gate.

\section{Device Fabrication, Experimental Setup, and Device Parameters}

Our experimental device consists of three flux-tunable Xmon qubits ($Q_1, C, Q_2$) with the middle one $C$ serving as the tunable coupler. Fabrication of this sample includes three main steps: (1) aluminum deposition onto a $c$-plane sapphire substrate followed by photolithography and inductively coupled-plasma etching to define all the base wiring and resonators; (2) two photolithography processes, aluminum deposition, and wet etching to construct airbridges~\cite{chen2014fabrication}; (3) e-beam lithography with two layer e-beam resists and double-angle aluminum deposition to make Josephson junctions. Airbridges are mainly used to connect segments of ground planes in order to reduce parasitic slotline modes. We also apply flux trapping holes (square holes of side length of 2~$\mu$m) to reduce magnetic vortices loss~\cite{Chiaro_2016}.

\begin{figure}[b]
\includegraphics{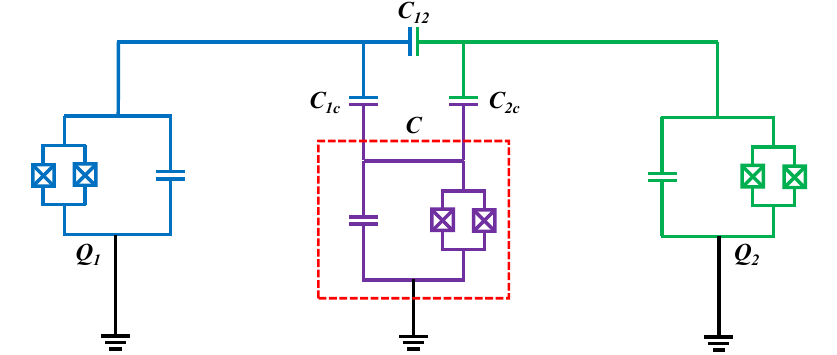}
\caption{Schematic electrical circuit of the device.}
\label{fig:DeviceSchematic}
\end{figure}

\begin{figure*}[hbt]
{\centering\includegraphics[width=1\textwidth]{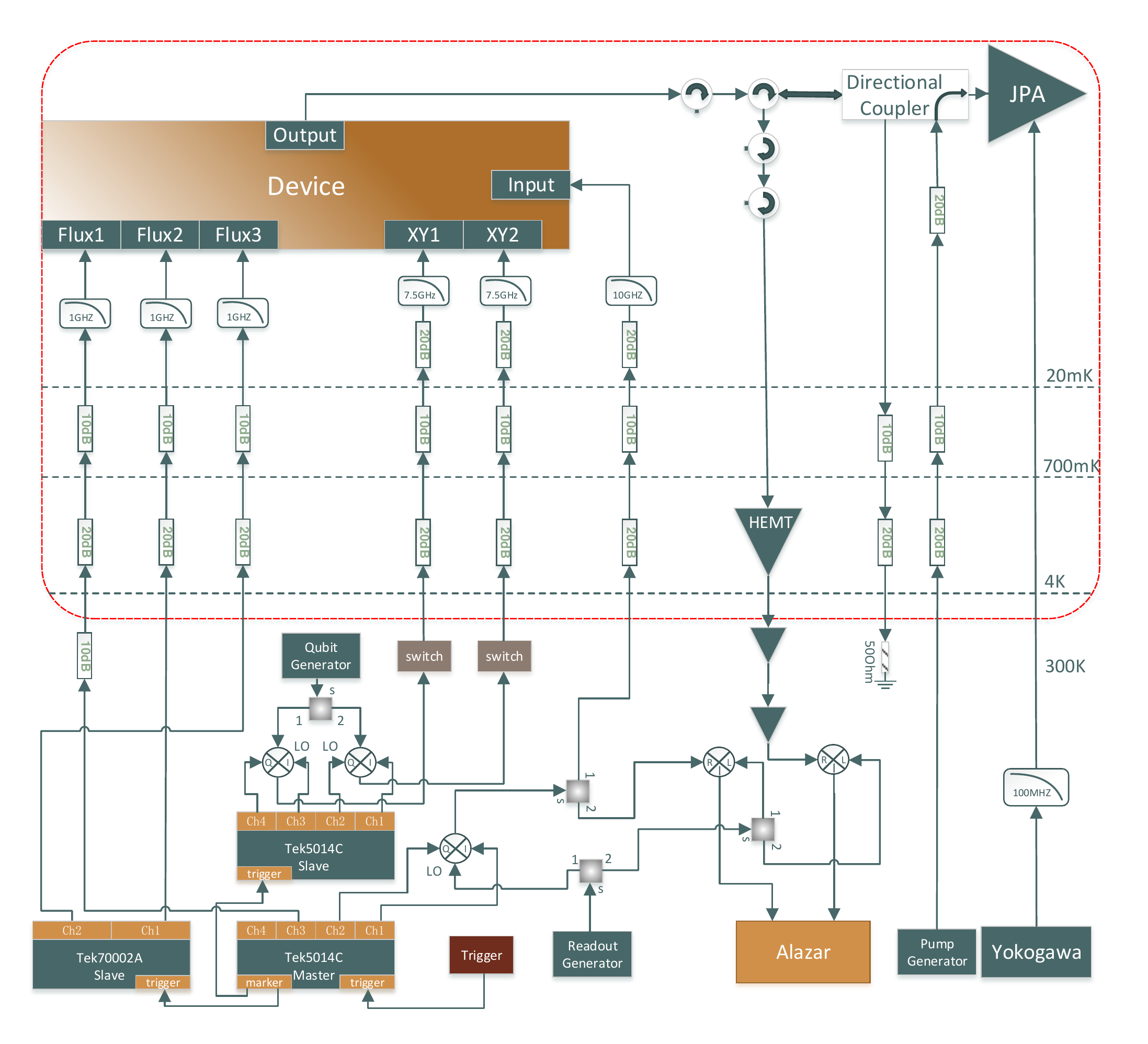}}
\caption{
Measurement setup. Our measurement circuit contains three AWGs (two Tek5014C and one Tek70002A), three signal generators, and other microwave components. The XY control pulses are generated from a signal generator modulated by the AWG. Flux pulses are generated directly from Tek5014C (for $Q_1$) and Tek70002A (for $Q_2$ and the coupler). Readout signal is amplified by a JPA at the base, a high-electron-mobility-transistor (HEMT) amplifier at 4K and two room-temperature amplifiers, and finally down-converted and digitized by an analog-to-digital converter (ADC).}
\label{fig:experimentsetup}
\end{figure*}

The full parameters of the qubits and the coupler are shown in Table~\ref{Table:parameters} with the coupling capacitances defined in Fig.~\ref{fig:DeviceSchematic}. Our sample is measured in a dilution refrigerator with a base temperature about 20~mK, and the details of our measurement circuit are shown in Fig.~\ref{fig:experimentsetup}. We use two XY control lines to manipulate the qubit states, three flux lines to modify the qubit and the coupler frequencies, and one input-output line to readout both qubits simultaneously. The XY control pulses are generated from a signal generator modulated by a four-channel arbitrary waveform generator (AWG), while flux pulses are directly generated from AWGs. Finally, a broadband Josephson parametric amplifier (JPA)~\cite{Hatridge,Roy2012,Kamal,Murch} is used for high-fidelity simultaneous single-shot readout.

Derivative removal adiabatic gate (DRAG)~\cite{motzoi2009simple} pulse is used for single qubit rotation to reduce the leakage to higher qubit levels. Due to thermal population of the qubits and the coupler, and nonperfect separation of the ground and excited states for each qubit, the qubit readout results are reconstructed by using Bayes' rule with a calibration matrix:
\begin{center}\label{S1}
${\rm M}_{{\rm B}j}$=
$\left(\begin{tabular}{cc}
${\rm F}_{{\rm g}j}$ & $1-{\rm F}_{{\rm e}j}$\\
$1-{\rm F}_{{\rm g}j}$ & ${\rm F}_{{\rm e}j}$\\
\end{tabular}\right)$\vspace{8pt},
\end{center}
where ${\rm F}_{{\rm g}j}$ and ${\rm F}_{{\rm e}j}$ are the readout fidelity for the $j$-th ($j=1,2$) qubit in the initial steady state without and with a following $\pi$ rotation respectively. The calibration process is similar to that in Ref.~\onlinecite{Li2018}.

Besides, we measure the flux-line-crosstalk matrix $M_z$ among the flux control lines (both qubits and the coupler) in the device. Although there is no readout cavity for the coupler, the coupler frequency can still be measured through the qubit-coupler dispersive shift (discussed below). The inverse of $M_z$ gives the orthogonalization matrix $\widetilde{M}_z$ which allows for independent control of the only desired qubit or the coupler:

\begin{center}
$\widetilde{M}_z$=$M_z^{-1}$=
$\left(\begin{tabular}{cccc}
0.9963 & 0.0096 & 0.0264  \\
$-0.0798$ & 0.9997  & 0.0094   \\
$-0.0116$ & 0.0384  & 0.9974  \\
\end{tabular}\right)$
\end{center}
The small flux-line-crosstalk is due to the good ground-plane connection by using airbridges even though the coupler is geometrically close to the two qubits.

\begin{table}[ht]
\caption{Device parameters.}
\begin{tabular}{cp{1.4cm}<{\centering}p{1.4cm}<{\centering}p{1.4cm}<{\centering}}
\hline
Qubit parameter &{$Q_1$}  &{$Q_2$} \tabularnewline
\hline
Readout resonator frequency (GHz) &{$6.825$} &{$6.864$} \tabularnewline
Qubit maximum frequency (GHz) &{$4.961$} &{$4.926$} \tabularnewline
$T_1$ (sweet point) ($\mu$s) &{$14$} &{$13.7$} \tabularnewline
$T_2$ (sweet point) ($\mu$s) &{$8.4$} &{$4$} \tabularnewline
$T_{2E}$ (sweet point) ($\mu$s) &{$8.7$} &{$4.4$} \tabularnewline
$\eta/2\pi$ (MHz) &{$-206$} &{$-202$} \tabularnewline
$\chi_{qr}/2\pi$ (MHz) &{$-0.4$} &{$-0.4$} \tabularnewline
$g_{qr}/2\pi$ (MHz) &{$86.6$} &{$90.6$} \tabularnewline
&&\tabularnewline
\hline
Coupler parameter &{Simulation}  &{Experiment} \tabularnewline
\hline
$\eta_c/2\pi$ (MHz) &{$-254$} &{} \tabularnewline
$C_{ic} (i=1,2)$ (fF) &{$2.4$} &{} \tabularnewline
$C_{12}$ (fF)&{$0.13$} &{} \tabularnewline
Coupler frequency (GHz) &{$6.3$} &{$5.977$} \tabularnewline

$g_{ic}/2\pi$ (i=1,2) (MHz) &{$81.3$} &{76.9} \tabularnewline
$g_{12}/2\pi$ (MHz) &{$3.8$} &{6.74} \tabularnewline
\hline
\end{tabular} \vspace{-6pt}
\label{Table:parameters}
\end{table}

\section{More Measurement Results}
\subsection{Coupler spectrum and qubit-coupler coupling strength}
\begin{figure}
\includegraphics{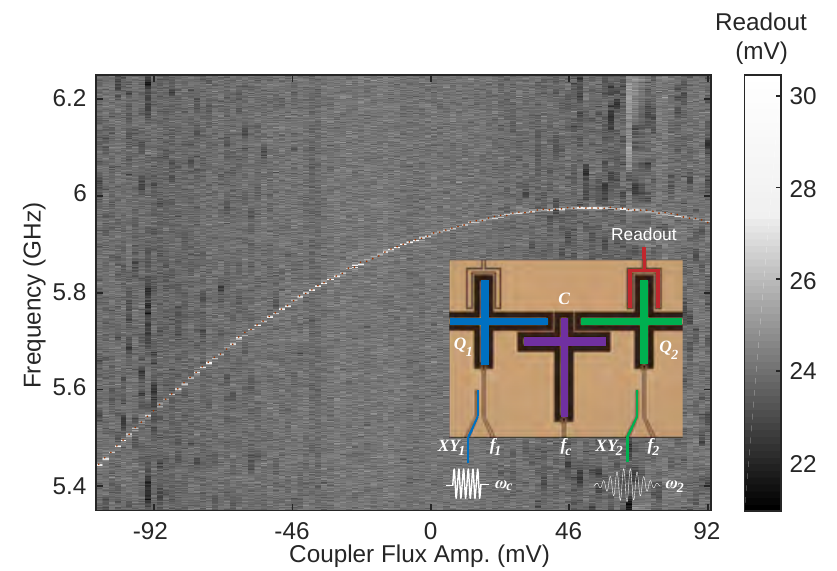}
\caption{Measured coupler spectrum. We drive the coupler through the XY control line of $Q_1$ by applying a microwave pulse at variable frequency with a rectangular envelope of a duration of 500~ns, followed by a population measurement of $Q_2$ with a wide selective Gaussian pulse. Inset: schematic of the measurement. The two computational qubits ($Q_1$ and $Q_2$) are shown in blue and green respectively, and the coupler is shown in purple.}
\label{fig:CouplerSpectrum}
\end{figure}

In our experiments, due to the lack of the readout resonator for the coupler, we cannot directly detect the coupler's thermal population and spectrum. Here, we just simply assume a similar thermal population for both the coupler and the computational qubits. We can, however, indirectly probe the coupler spectrum via the qubit-coupler dispersive shift, by driving the coupler with the XY control line of $Q_1$ followed by a population measurement of $Q_2$. The coupling between each qubit and the coupler causes a dispersive frequency shift when they are far detuned as:
\begin{equation}\label{Eq13}
\begin{split}
\chi_{ic}=\frac{(g_{ic})^2(\eta_i+\eta_c)}{2(\Delta_{ic}-\eta_c)(\Delta_{ic}+\eta_i)}.
\end{split}
\end{equation}
The experimental results of coupler spectrum are shown in Fig.~\ref{fig:CouplerSpectrum}.

The qubit-coupler dispersive shift may also decrease the single-qubit gate fidelity when the coupler has a large thermal population. However, in our experiments, the single-qubit gate is implemented when the frequency detuning is large between the qubits and the coupler. Thus, the coupler's thermal population would have little effect on the single-qubit gate fidelity and overall system preparation. In addition, we believe, through carefully designing our sample and fridge environment, the coupler thermal population can be suppressed to a very low level. 

\begin{figure}
\includegraphics{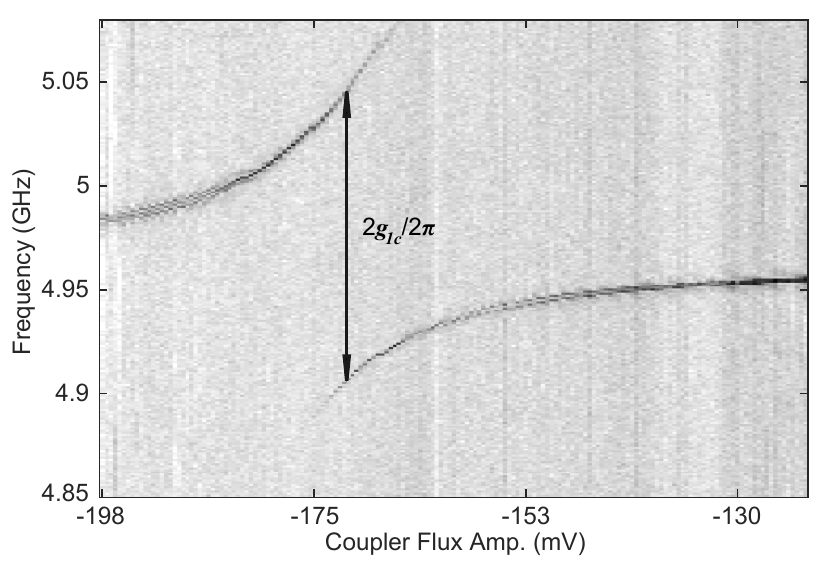}
\caption{Qubit-coupler anti-crossing. At the resonance point, anti-crossing of the energy level is observed and the separation characterizes the coupling strength.}
\label{fig:QubitCouplerAntiCrossing}
\end{figure}

\begin{figure*}
\includegraphics{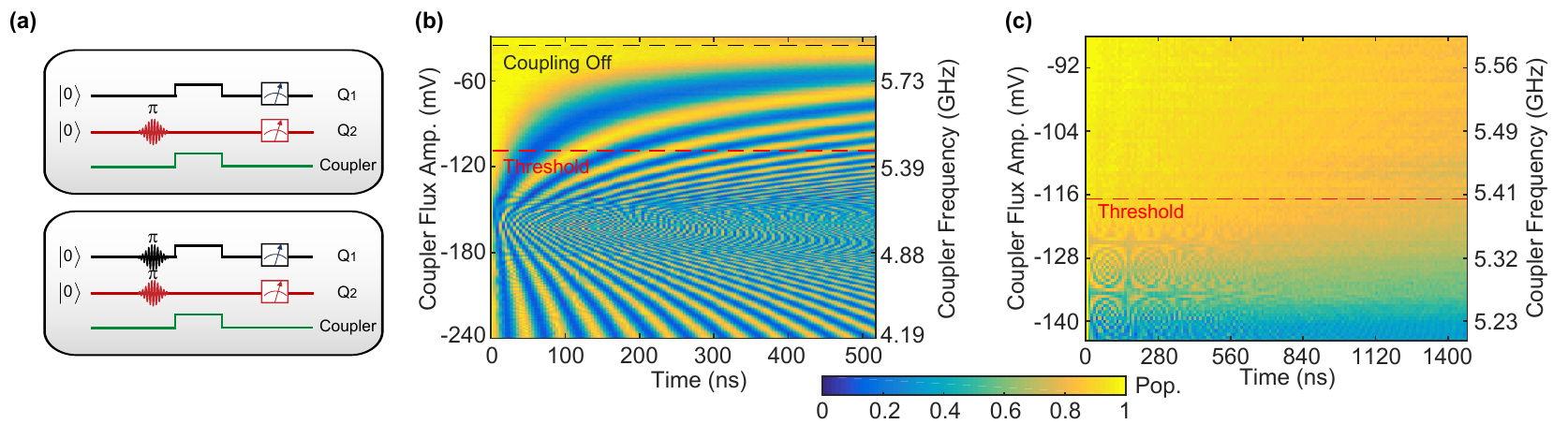}
\caption{Lower threshold for the coupler frequency. (a) The pulse sequences to measure the operation range with negative coupling strength. (b) A zoomed-in part of Fig.~\ref{fig:Fig2}(b) in the main text. The swap interaction is completely turned off at the marked dark dashed line, while the red dashed line indicates the approximate threshold for non-negligible excitation of the coupler. (c) Direct leakage experiment for finding the approximate threshold of the coupler frequency. The two qubits are initialized in $\ket{11}$, then are tuned into resonance while varying the coupler frequency, and finally the population of $\ket{11}$ state is measured. The red dashed line shows the threshold of having leakage out of the computational space.}
\label{fig:CouplerThreshold}
\end{figure*}

The coupling strength between each qubit and the coupler can be measured in a swap operation between them. We perform the qubit spectroscopy measurement while varying the coupler frequency and biasing the other qubit far away. The anti-crossing in the spectrum can be seen as shown in Fig.~\ref{fig:QubitCouplerAntiCrossing} and  the qubit-coupler coupling strength $g_{1c}$ can be extracted.

\subsection{Operation range with negative coupling strength}
To perform the iSWAP and $\rm \sqrt{iSWAP}$ gates, we need to choose an appropriate coupling strength by varying the coupler frequency when the two qubits are tuned into resonance. To tune the qubit and the coupler, we use fast rectangular flux pulses for the measurements shown in Fig.~\ref{fig:Fig3} in the main text.

In order to achieve fast two-qubit gate operations, the coupling strength needs to be large. The positive coupling strength is defined and limited by the geometry of the device, while the negative value can be varied by tuning the coupler frequency. However, the coupler frequency cannot be tuned too closed to the qubit frequency for a large negative coupling strength without causing direct energy exchange between the qubit and the coupler, i.e. the leakage out of the computational space. Figure~\ref{fig:CouplerThreshold}(b) is a zoom-in and finer sweep of  Fig.~\ref{fig:Fig2}(b) of the main text. The red dashed line shows an approximate threshold, a lower threshold for the coupler frequency, below which small ripples in the population oscillation of $Q_2$ start to show up, indicating non-negligible excitation of the coupler. The leakage out of the computational space can also be directly measured by monitoring the population of $\ket{11}$ state when both qubits are excited. The experimental results are shown in Fig.~\ref{fig:CouplerThreshold}(c). Again, below the threshold of the coupler frequency, the population of $\ket{11}$ state starts to deviate from the desired exponential decay. The experimental sequences for the measurements performed in Figs.~\ref{fig:CouplerThreshold}(b) and \ref{fig:CouplerThreshold}(c) are shown in Fig.~\ref{fig:CouplerThreshold}(a).

\begin{figure}
\includegraphics{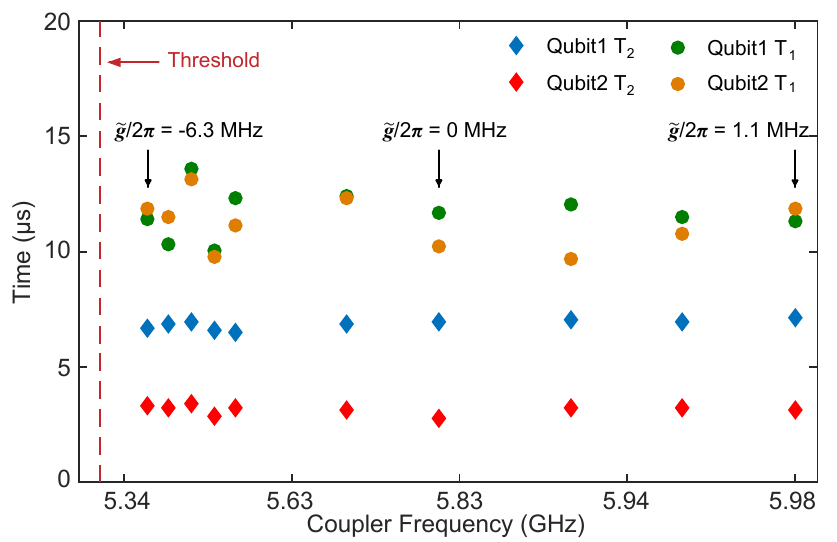}
\caption{Qubit coherence times vs coupler frequency. When the coupler frequency is above the threshold for non-negligible excitation of the coupler, the qubit coherence times remain nearly unaffected. Three specific coupling strengths are marked by the black arrows.}
\label{fig:QubitCoherence}
\end{figure}

\subsection{Qubit coherence}

One of the advantages of this prototype of coupler is the flexibility and compatibility to a large-scale architecture. The positive coupling strength is defined by the direct coupling between the two Xmon qubits, while the negative coupling is provided by the coupler. The competition between the positive direct and negative indirect coupling allows for a continuous tunability and for switching off the coupling completely. The change of sign of the coupling strength could also provide a valuable degree of freedom for future quantum simulation applications.

Besides, this coupler scheme can also have little impact on the qubit coherence through careful design, although the coupler is capacitively coupled to both qubits and may offer an additional decay channel to the qubits. In addition, the coupler flux bias line may inductively couple to the qubits and affect their coherence. We measure the qubit coherence times at different coupler frequencies, as shown in Fig.~\ref{fig:QubitCoherence}. In the operation range above the threshold, the qubit coherence times remain unaffected regardless of the coupling strength.

\begin{figure}
\includegraphics{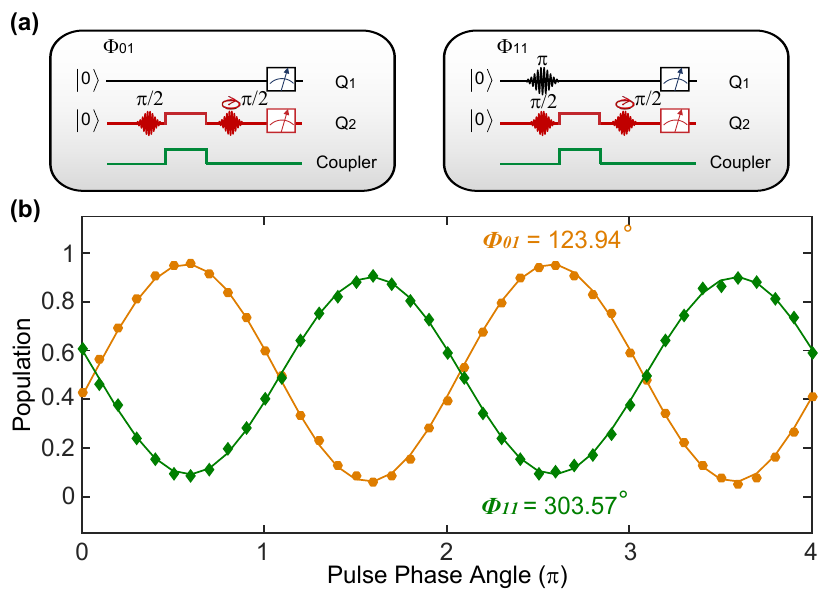}
\caption{The CZ gate calibration. (a) The experimental sequences to measure $\phi_{10}$, phase accumulation of $\ket{01}$ with respect to $\ket{00}$, and $\phi_{11}$, phase accumulation of $\ket{11}$ with respect to $\ket{10}$. (b) The probabilities of measuring $\ket{01}$ and $\ket{11}$ vs the phase of the second $\pi/2$ pulse. The sinusoidal oscillations reveal the phases acquired in the Ramsey measurements for the two-qubit states $\ket{01}\rightarrow e^{i\phi_{01}}\ket{01}$ and $\ket{11}\rightarrow e^{i\phi_{11}}\ket{11}$, respectively. The solid lines are fits to sinusoidal oscillations. With a standard CZ gate time of $t_R$=222~ns, a $\pi$ phase shift between the orange and green curves is observed.}
\label{fig:CZcalibration}
\end{figure}

\subsection{Calibration of the CZ gate}
We first show the calibration of the standard CZ gate implemented by using rectangular flux pulses. We conduct the Ramsey measurements to extract the single qubit phase $\phi_{01}$ and the two-qubit conditional phase $\phi_{11}$, as shown in Fig.~\ref{fig:CZcalibration}(b). The solid lines are fits to sinusoidal oscillations. A $\pi$ phase shift between $\phi_{01}$ and $\phi_{11}$ curves is observed when both qubits are excited.  We correct the single-qubit phases in software to acquire the desired CZ gate with the unitary matrix as diag$\{1, 1, 1,-1\}$. The CZ gate fidelity is estimated to be 95.5\% on average from the QPT measurement. We use the same method to calibrate the conditional $\pi$ phase of the CZ gate with the dynamically decoupled regime (DDR) mentioned in the main text.

\section{Simulation of the CZ gate}

\begin{figure}
\includegraphics{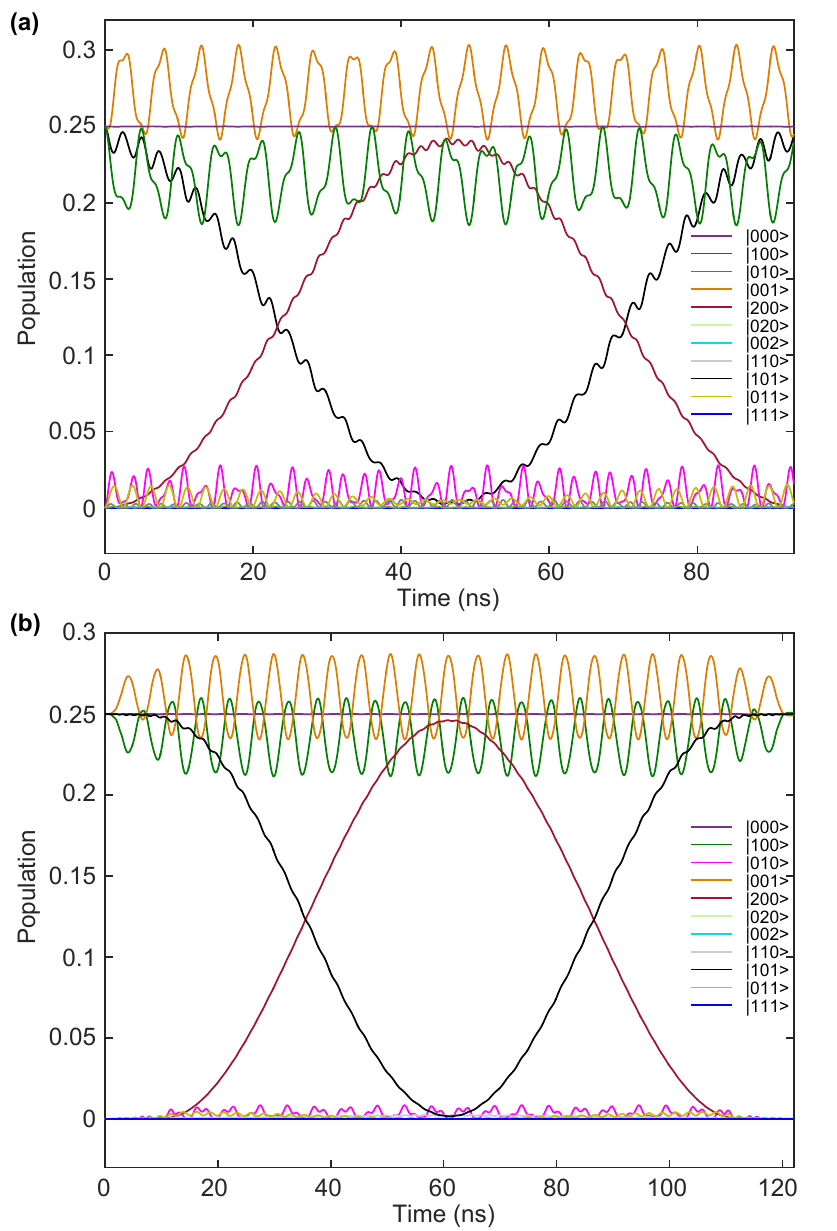}
\caption{Simulation of the population evolution of each state during the CZ gate with an initial joint qubit-coupler state $\ket{Q_1CQ_2}=\ket{+0+}$. (a) CZ gate with rectangular flux pulses. (b) CZ gate with DDR.}
\label{fig:CZsimulation}
\end{figure}

We simulate the CZ gates with QuTip in Python~\cite{johansson2012,johansson2013}. First, we confirm the coupler decoherence has little effect on the CZ gates. The coupler frequency can be tuned in a large range. If the coupler frequency is far detuned from its sweet spot, the dephasing of the coupler will increase significantly. We compare the CZ gate fidelities for two cases in which the coupler dephasing time $T_2= 5~\mu$s and $T_2=0.5~\mu$s, and find there is nearly no difference between them. This is because the coupler remains almost on the ground state for the whole process.

We then simulate the standard CZ gate implemented by the rectangular flux pulse (with `mesolve' in QuTip). The population evolution of each state with an initial state $\ket{Q_1CQ_2}=\ket{+0+}$ is shown in Fig.~\ref{fig:CZsimulation}(a), where $\ket{+}=(\ket{g}+\ket{e})/\sqrt{2}$. In this case, the coupler frequency is tuned to 5.337~GHz and the total qubit-qubit coupling is negative. We observe significant high-frequency oscillation between $\ket{101}$ and $\ket{200}$. This non-ideality comes from the population exchange between the coupler and the qubit, evidenced by the oscillation between $\ket{010}$ and $\ket{100}$. The swap oscillation between $\ket{001}$ and $\ket{100}$ is owing to the anharmonicity of the qubits. After a synchronization optimization strategy~\cite{Barends2019}, we can achieve the fidelity of this type of CZ gate $F=98.75\%$ in 90~ns with no system coherence.

As a comparison, the population evolution of each state for the CZ gate with the DDR scheme with the same initial state $\ket{Q_1CQ_2}=\ket{+0+}$ and with the same solving methods is shown in Fig.~\ref{fig:CZsimulation}(b). In this simulation, the gate time is 120~ns including DDR. The oscillation amplitude between the coupler and $Q_1$ decreases significantly when the coupler is turned on. If we use a slower cosine rising edge, the oscillation amplitude gets even smaller, but at the cost of a longer operation time. Most importantly, when we slowly turn off the coupler by tuning up the coupler frequency, the swap oscillation between the coupler and $Q_1$ gets much weaker, so the population of the coupler would slowly oscillate back into the computational space. 

Based on simulation, this CZ gate fidelity could be as high as $F=99.998\%$ after careful synchronization optimization strategy~\cite{Barends2019}, provided there is no system decoherence. Given the experimental coherence times, a simulated CZ gate fidelity $F=98.1\%$ is acquired, in good agreement with the measured $F=98.3\pm0.6\%$ in the main text. In fact, such variations of the experimental gate fidelity are mainly due to two reasons. First, in our experiment, the measured histogram threshold between the ground state and the excited state of each qubit could fluctuate with time due to low frequency drift in the measurement setup. Following the Bayes rule calibration method (details can be found in Appendix C), we would acquire a measurement fluctuation distribution, and thus a variation of the gate fidelity. Second, the coherence of each qubit also varies with time. On one hand, this can affect the gate performance and thus lead to a fluctuating CZ gate fidelity; on the other hand, the simulated gate fidelity sensitively depends on the coherence times we use. Our simulation shows that the CZ gate fidelity can be improved from F = 98.1\% to F = 98.3\% by increasing $T_2$ of each qubit by only about 10\%. 

In addition, we also follow the method developed in Ref.~\onlinecite{korotkov2013} to extract error sources in our experimental CZ gate. We find that the decoherence error is the main error source with a contribution of about 1.14\%. Other errors, contributing only 0.56\% in total, may come from the state preparation and measurement errors (details can be found in Appendix B).

Because of the unwanted transitions, we could not acquire a complete geometric $\pi$ phase for the CZ gate with DDR. The simulation shows the geometric contribution to the $\pi$ phase is about 98.3\%. We have to tune the operation point slightly away from the resonance between $\ket{101}$ and $\ket{200}$ to accumulate a small dynamical phase (about 3 degrees) to realize the required $\pi$ phase.

With the extra degree of freedom provided by the tunable coupler, a more efficient pulse shape could be optimized to achieve a CZ gate with a higher fidelity and lower unwanted leakage in future experiments. This deserves future exploration.

\normalem

%

\end{document}